# Targeted Resilient Zoning for High-Impact Events via Multi-Circuit Polelines

Hritik Gopal Shah, Gregory Giustino and Elli Ntakou, *Member, IEEE*[1]

*Abstract*—The increasing frequency and severity of High-Impact Low-Probability (HILP) events such as hurricanes and windstorms pose significant challenges to the resilience of electrical power distribution systems, particularly in regions of New England where there is a significant amount of overhead infrastructure in areas where vegetation is predominant. Traditional reliability-focused planning is insufficient to address the systemic vulnerabilities exposed by such extreme events. This paper presents a novel risk-based framework for long-term resilience planning of active overhead distribution systems, with a specific focus on mitigating the impacts of high-wind and hurricane-induced outages. The proposed methodology identifies and analyzes Multiple Circuit Pole (MCP) lines, these are critical infrastructure segments where multiple circuits share common support structures across the utility's service territory. Delineating MCP risk zones, quantifying their spatial extent, and cataloging associated attributes such as pole IDs, circuit composition, and zone lengths enables targeted hardening strategies and supports data-driven decisions to enhance grid resilience.

## I. Introduction and Motivation

Extreme weather conditions, often referred to as High-Impact Low-Probability (HILP) events [1], pose a growing threat to the operational resilience of modern power systems. These events, such as hurricanes, floods, wildfires, and ice storms, are becoming more frequent and severe due to climate change. The Intergovernmental Panel on Climate Change (IPCC) has projected that global warming will continue to intensify the frequency and magnitude of such extreme events, with significant implications for critical infrastructure [2].

Power systems, particularly those with extensive overhead infrastructure, are exposed to weather-caused disruptions. Severe weather events have already demonstrated their capacity to cause widespread outages and cascading failures across interconnected grid components. In the United States alone, weather-related disruptions to transmission and distribution networks result in annual economic losses estimated at average $100 billion [3]. Hurricanes, in particular, are responsible for the majority of long-duration outages, often lasting several days and affecting millions of customers. Globally, similar patterns are observed.

For example, Typhoon Jebi in Japan (2018) [4] caused extensive damage to power infrastructure, Tropical Storm Isaias impacted the whole New England region leading to prolonged outages and significant economic losses [5]. In a survey of 500 households in Houston [6], Hurricane Harvey in 2017 resulted in nearly 70% of respondents experiencing power outages averaging 5 days, with the longest lasting over 300 days.

These events highlight the systemic vulnerability of power systems to HILP events. Tree-related incidents are a leading cause of power outages during storms based on Eversource's outage data, accounting for over 90% of service interruptions in some events. The risk is further amplified by the high percentage of tree canopy over distribution lines. According to LiDAR-based studies and USDA Forest Service datasets, tree canopy coverage in New England can exceed 60% in urban areas and is even higher in suburban and rural zones. This extensive canopy, while ecologically valuable, increases the likelihood of vegetation interference with overhead lines. Climate change is expected to exacerbate these challenges by negatively impacting tree health through increased pest infestations, drought stress, and more frequent extreme weather, thereby increasing the likelihood of tree failures and outages.

The challenges posed by tree coverage, system vulnerabilities and HILP events underscore the urgent need for advanced resilience assessment and planning approaches that extend beyond conventional reliability and resiliency metrics [7]. While existing research primarily focuses on identifying critical transmission lines or branches using probabilistic failure models, optimization techniques and proposing hardening strategies [8][9], our work, by contrast, introduces a data-driven methodology for large-scale utility service territory, that targets a previously underexplored vulnerability: multi-circuit pole lines (MCPs).

---

[1] Hritik Gopal Shah, Gregory Giustino and Elli Ntakou are with Reliability and Resiliency Department in Eversource Energy, USA (email: hritik.shah@eversource.com, gregory.giustino@eversource.com and elli.ntakou@eversource.com)

A significant vulnerability emerges when multiple electrical circuits are supported by shared pole infrastructure, commonly referred to as Multiple Circuit Poles (MCPs) [11]. Damage to these MCPs can result in common mode failure (CMF) events, leading to extensive service disruptions and operational complexities [12]. These events are particularly critical when the circuits are designed to provide mutual redundant support under N-1 contingency scenarios, as a single point of failure can simultaneously compromise multiple backup paths[8]. Observations during major storm events have shown that MCP locations are among the most impactful, often resulting in large-scale outages due to the simultaneous loss of primary service and the unavailability of alternate circuit ties.

Traditional measures and historical data often fail to capture the systemic risks posed by common mode failures and the potential loss of redundancy in highly meshed interconnected distribution systems for a large electric utility operating across multiple operating regions or states. As this has thousands of miles of overhead (OH) distribution and transmission lines and manually identifying and managing multi-circuit pole lines is not feasible and currently, no standardized process exists for doing so. An automated process is essential to efficiently map, monitor, and prioritize these locations. This paper proposes a risk-based framework for identifying and analyzing MCP risk zones across the service territory. The paper makes the following key contributions to the field of electric distribution system resiliency and spatial infrastructure analytics.

- Developed novel spatial analytics algorithms based on proximity clustering to automatically identify all properties of MCP lines, including the circuits involved and the exposure length of the common runs of the circuits on the MCPs. These analytics are useful to utilities to bring together disjoint data to operationalize and act on data from multiple sources

- Designed a multi-criteria prioritization framework to rank MCPs for targeted resiliency upgrades and undergrounding based on system impact and vulnerability.

## II. METHODOLOGY

### A. Dataset Description

The analysis is based on Geographic Information System (GIS) infrastructure data of Eversource Energy, covering its entire service territory. The dataset comprises two primary components.

- Pole Data: Denoted as set $P = \{P_1, P_2, P_3, \ldots, P_n\}$ where each pole $P_i$ contains attributes such as unique Pole ID, geographic coordinates $X_i = (x_i, y_i)$ and structural metadata.
- Overhead Wire Segments: Represented as set $W = \{W_1, W_2, W_3, \ldots, W_m\}$, where each segment $W_j$ is a discrete line element with associated attributes Length $L_j$, Circuit ID $C_j$, Ampacity or capacity $A_j$, Pole IDs $P_i$.

It is important to note that the overhead wire data $W$ is not topologically continuous in the dataset; segments are not explicitly connected in terms of data structure. However, spatial continuity is implied through GIS geometry and segments are physically adjacent or "touching" in the spatial domain.

### B. Multiple Circuit Polelines Risk Zone identification

The MCP (Multiple Circuit Pole) identification and clustering process has been automated using ArcGIS–Python integration and applied across the three-state service territory of Eversource Energy. The methodology consists of three main stages.

1. Pole-to-Overhead Wire Segments Association

In the Eversource Energy GIS infrastructure, there is no direct one-to-one data linkage between poles and overhead wire segments. This is due to spatial offsets in the GIS data, where overhead wire geometries may not be co-located with pole geometries. While this lack of direct linkage between poles and overhead wire segments is a known issue within Eversource Energy's GIS infrastructure. It is not universal across all utilities, depending on their GIS data modeling practices and legacy system constraints.

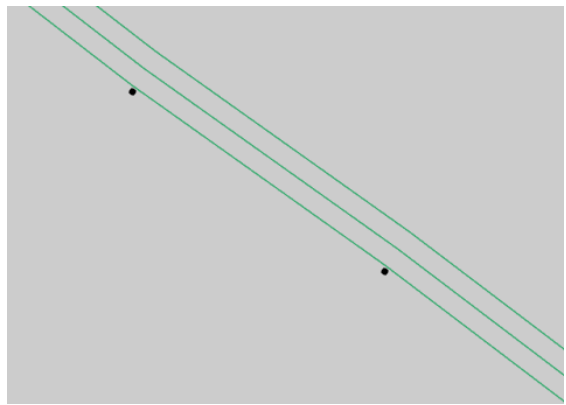

Fig 1: Actual GIS data structure of Poles (blue dots) and Overhead wire segments (green lines)

To infer the most likely overhead wire associations for each pole, we apply a k-Nearest Neighbors (k-NN) [13] model as a spatial proximity filter. In spatial analysis, proximity is a strong indicator of connectivity, making k-NN an effective tool for estimating pole-to-wire relationships. The lateral displacement between a pole and its associated overhead wire is varying across the service territory so we use the Euclidean distance (1) between pole $P_i$ with coordinates $X_i = (x_i, y_i)$ and wire segment centroids $W_j = (x_j, y_j)$

$$d(X_i, W_j) = \sqrt{(x_i - x_j)^2 + (y_i - y_j)^2} \quad (1)$$

By analyzing dataset, no pole in the Eversource system carries more than three overhead circuits and not more than 50 meters away from the pole. Thus, we set $k = 3$ to ensure the model considers all potential circuit associations and $d \leq 50$.

$$N_3^{(r)}(P_i) = \{W_j : d(X_i, W_j) \leq 50 \ meters\} \quad (2)$$

(2) ensures that only wires within a realistic spatial proximity are considered, reducing false associations. These inferred associations are then used in downstream MCP identification and risk zone analysis.

2. GIS-Based Multi-Circuit Pole Detection

Now as we have associated each pole $P_i$ with overhead segment wires $W_j$, each pole is thus defined by tuple $P_i = (W_j, X_i, C_j)$. Based on our nomenclature, the third parameter of the tuple, the circuit names, is inferred through the conductor segment $W_j$. A pole $P_i$ is classified as a multi-circuit pole if it carries more than 1 unique circuits (3).

$$|\{C_j : P_i\}| > 1 \quad (3)$$

3. K-Dimensional Tree-Based MCP Risk Zone Clustering Algorithm

Having identified poles that support multiple unique circuits, the next step involves spatially clustering these poles based on both their shared circuit configurations and geographic proximity into risk zone $Z_k$. These clusters represent critical infrastructure segments where multiple circuits are co-located on the same set of poles within a confined area. Such configurations represent critical points of vulnerability during HILP events, such as severe storms or localized infrastructure failures and damage to any pole within these clusters could simultaneously disrupt multiple circuits and lead to disproportionately large customer outages. By delineating these risk zones and calculating their spatial extent. This enables utilities a data-driven approach to resilience planning, ensuring that infrastructure investments are focused on the most vulnerable and high-impact segments of the overhead network for optimize targeted undergrounding strategies. This section presents a novel KD-Tree-based algorithm for efficient clustering and spatial extent estimation of MCP risk zones.

Let $P_{MCP_i} = \{P_{MCP_1}, P_{MCP_2}, \dots, P_{MCP_k}\}$ be a finite set of multi-circuit pole locations in $\mathbb{R}^2$, where each multi-circuit pole $P_{MCP_i}$ has coordinates $X_i$. Let $r$ be a fixed spatial proximity threshold. Then, the following algorithm partitions $P_{MCP}$ into disjoint spatial clusters $Z_1, Z_2, \dots, Z_k$, each representing a risk zone. To enable efficient spatial queries, a KD-Tree $T$ is constructed from the set of MCP coordinates $\chi = \{X_1, \dots, X_n\}$.

The KD-Tree recursively partitions the 2D space by alternating between the $x$ and $y$ axes at each level of the tree. The construction has a time complexity of $O(n \log n)$.

Next, a Radius-Based Neighborhood Query is applied (4). For each unvisited MCP $X_i \in \chi$, a fixed-radius neighborhood query is performed using the KD-Tree to identify all MCPs within a specified distance threshold $r$. The neighborhood set is defined as

$$N_r(X_i) = \{X_j \in \chi \mid \|X_j - X_i\|_2 \leq r\} \quad (4)$$

where $\|.\|_2$ denotes the Euclidean norm, and $X_j$ is a coordinate in $\chi$ that lies within radius $r$ of $X_i$. The value of $r = 200$ meters are selected based on engineering judgment and spatial density analysis of poles in Eversource Energy territory. It basically means that the maximum distance between adjacent poles is approximately 200 meters.

---

**Algorithm 1 MCP Risk Zone Clustering Using KD-Tree**

---

**Input:** Set of MCP coordinates: $\chi = \{X_1, \dots, X_n\}, x_i \in \mathbb{R}^2$, Proximity threshold: $r > 0$ (e.g., 200 meters) and pole $P_i$ with same multi-circuit configuration.

**Output:** Set of disjoint clusters: $Z = \{Z_1, Z_2, \dots, Z_k\}$ and estimated spatial extent of each risk zone $D_k$

**Procedure**:
1. Construct a KD-Tree $T$ from the input set $\chi$ to enable efficient spatial queries.
2. Initialize: Let $Z \leftarrow \emptyset$ (set of clusters) and $V \leftarrow \emptyset$ (set of visited points)
3. For each unvisited pole $P_i \in \chi \setminus V$:
- Initialize a new cluster $Z_k = \{X_i\}$.
- Add $X_i$ to $V$.
- Initialize a queue: $Q \leftarrow \{X_i\}$.
4. While $Q \neq \emptyset$:
- Dequeue a point $x_q$ from $Q$
- Query KD-Tree for neighbors: $N_r(X_i) = \{X_j \in \chi \mid \|X_j - X_i\|_2 \leq r\}$
- For each $X_j \in N_r(X_i) \setminus V$:
  - Add $X_j$ to $Z_k$
  - Add $X_j$ to $V$
  - Enqueue $X_j$ into $Q$
5. Add cluster $Z_k$ to $Z$
6. Repeat steps 3–5 until all points in $\chi$ are visited and assigned to a cluster.
7. For each cluster $Z_k = \{X_{i_1}, \dots, X_{i_m}\}$, estimate its spatial extent: $D_k = \sum_{j=1}^{m-1} \|x_{i_{j+1}} - x_{i_j}\|_2$. This gives the total pole-to-pole distance within the cluster

---

## III. RESULTS

From the above proposed methodology a total of 1,041 MCP risk zone locations were identified across Eversource Energy's tri-state service territory. Notably, all identified MCP risk zones exceeded 200 meters in length, indicating substantial spatial exposure during HILP events. To better understand the spatial extent of these zones, a distribution analysis was conducted across all three states in the service territory. As shown in Fig 2, there are 824 MCP risk zones ranging from 0.2 to 1.2 miles in length, along with 24 zones exceeding 5 miles. This wide variation highlights the importance of spatial clustering in identifying extended risk corridors. Fig 3 shows a particular MCP risk zone identified by our proposed methodology in our Connecticut service territory.

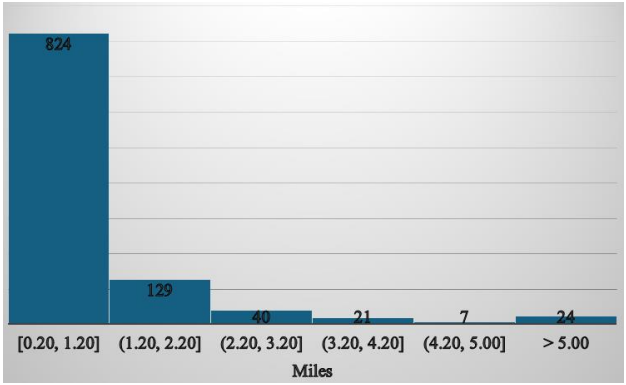

Fig 2. Distribution of MCP risk zone count across distance ranges with 1-mile binning across Eversource Energy service territory.

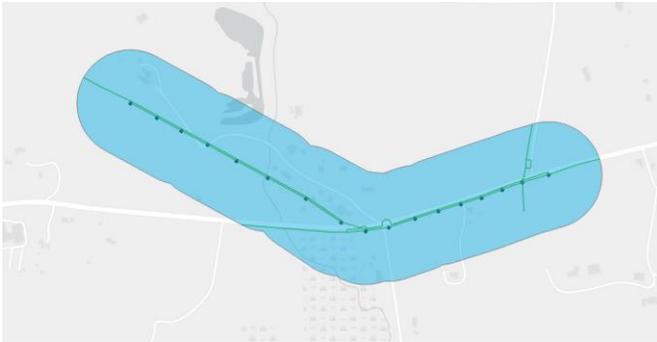

(a)

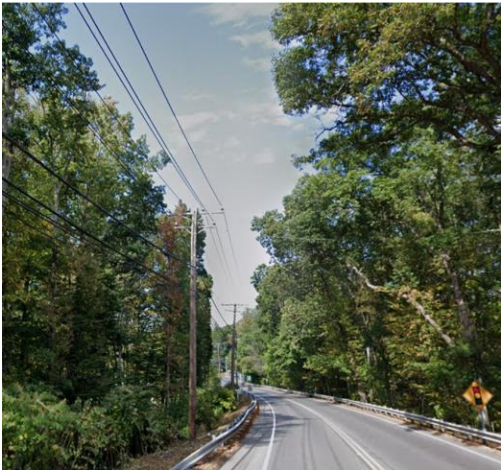

(b)

Fig 3: (a) A MCP Risk zone identified with Poles (blues dots), Overhead Power lines (green lines) and MCP risk zone (light green polygon). (b) Same MCP risk zone in (a) visualized on Google Street View [15], highlighting the multi-circuit pole lines.

A focused analysis of historical outage data from four representative MCP locations identified from our process underscores the operational vulnerabilities associated with multi-circuit configurations. During Tropical Storm Isaias, these MCP locations were among the worst-performing segments of the distribution system. Table I summarizes the outage impact metrics for these locations, characteristics like Right-Of-Way (ROW) or roadside, construction type and number of circuits involved.

Table I: Historical Outage Analysis of 4 MCP risk zone locations

| Location Type | No of Circuits | Customers Affected | Customer Minutes Interrupted (CMI) | Construction Type |
|---|---|---|---|---|
| Roadside | 3 | 4,764 | 22,027,105 | Circuits 1 & 2 vertically stacked on crossarms with covered conductor; Circuit 3 aerial cable below |
| Roadside | 3 | 3,913 | 29,293,633 | Circuit 1 covered conductor on crossarm; Circuits 2 & 3 aerial cable below |
| ROW | 2 | 9,828 | 29,208,816 | Circuits 1 & 2 horizontal construction on the same crossarm, both bare conductor |
| ROW | 2 | 12,412 | 21,781,596 | Circuits 1 & 2 vertically stacked on crossarms with bare conductor |

While the average customer outage during Isaias affected approximately 630 customers with 1.78 million CMI, MCP-related events impacted 3,900 to 12,400 customers each, with CMI values ranging from 21 to 29 million. All four locations involved mutually dependent circuits, where a single point of failure disrupted multiple restoration paths. This compounded outage severity and extended restoration timelines. Notably, both roadside and right-of-way (ROW) configurations were affected, regardless of whether they employed covered or bare wire. These findings suggest that during severe weather events, construction type alone does not mitigate outage risks, especially the systemic risk posed by MCPs.

To support targeted infrastructure hardening, a multi-factor prioritization framework was developed to rank MCPs for resiliency upgrades. Each MCP was scored based on seven weighted criteria, as shown in Table II.

Table II: Prioritization framework to rank MCP zones for targeted resilience planning

| Factor | Description | Weight (%) |
|---|---|---|
| Customer Impact | Number of customers affected by MCP failure | 30% |
| Redundancy Loss Severity | Degree to which MCP failure compromises backup/tie paths | 20% |
| Critical Infrastructure | Presence of hospitals, emergency services, or essential facilities | 10% |

| Asset Condition | Health of poles, conductors, and associated equipment | 20% |
| Restoration Complexity | Difficulty of access and repair, especially in ROW areas | 10% |
| Outage Risk | Physical vulnerability (e.g., bare wire, vegetation exposure) | 10% |

This scoring model enables a data-driven approach to prioritize MCPs that pose the greatest risk to system reliability and customer service continuity. MCPs with high customer impact, redundancy loss, and poor asset condition were ranked highest for reinforcement.

## IV. Conclusion

This work presents a novel automated framework for resiliency planning across an electric service territory during high-impact, low-probability (HILP) events. Our approach leverages real-world infrastructure and outage data to uncover spatial clusters of MCPs that pose a high risk of common mode failures. By integrating spatial clustering techniques with circuit configuration analysis, we delineate risk zones that are both geographically coherent and operationally critical. This enables the development of targeted hardening strategies such as selective undergrounding or structural reinforcement prioritized by actual exposure and impact potential. Our methodology not only enhances the granularity and relevance of resilience planning but also provides a scalable tool for utilities seeking to mitigate cascading outages in the face of increasingly severe climate-driven disruptions.